# Evoking Places from Spaces: The application of multimodal narrative techniques in the creation of "U - Modified"

**G. W. Young**
National Centre for Geocomputation
Maynooth University
Maynooth, Ireland.
`gareth.young@mu.ie`

**S. Mannion**
Department of Music
University College Cork
Cork, Ireland.
`Siobhannimhainnin@gmail.com`

**S. Wentworth**
Department of Music
University College Cork
Cork, Ireland.
`sara@sarawentworth.com`

## ABSTRACT

Multimodal diegetic narrative tools, as applied in multimedia arts practices, possess the ability to cross the spaces that exist between the physical world and the imaginary. Within this paper we present the findings of a multidiscipline practice-based research project that explored the potential of an audio-visual art performance to purposefully interact with an audience's perception of narrative place. To achieve this goal, research was undertaken to investigate the function of multimodal diegetic practices as applied in the context of a sonic-art narrative. This project direction was undertaken to facilitate the transformation of previous experiences of place through the creative amalgamation and presentation of collected audio and visual footage from real-world spaces. Through the presentation of multimedia relating to familiar geographical spatial features, the audience were affected to evoke memories of place and to construct and manipulate their own narrative.

## 1. INTRODUCTION

The use of diegesis as a narrative tool is typically applied via music or other types of sound effects in support of a storyline delivered through a visual medium, a topic often explored in contemporary media studies. This theory can be further explored in multimedia art to effectively summarise events that occur in a performance and can serve to add further commentary on the intentions and thoughts of the artist. By applying this narrative tool, it was hypothesised by the collaborating artists on the "U - Modified" project that memories of place could be evoked from within an audience as they experience an artistic multimedia representation of geographic and narrative spaces.

To create an affected sense of place, the project explored the use of real-world sounds and site-specific video footage to inform and manipulate the mental construction of narrative. Through the multimodal presentation of these real-world spaces, a performance was devised that conjured brief tangential storylines that were to be interpreted by the audience to construct an internalised responsive narrative. It was therefore in the process of interpretation that the narrative place was to be moulded; specifically, by transforming the performance materials into memories of previous experiences and for the audience to then embody the created narrative. The audiences' knowledge of the material spaces they were presented with were therefore to be manipulated to explore the use of a responsive storyline, as perceived through a multimedia performance.

To explore the use of diegesis in this context, Young's original composition of "U" (2013) was modified with the addition of extended vocal techniques and visual footage, provided by Mannion and Wentworth respectively. For each of these modifying elements, vocal features were scored, arranged, and performed by Mannion and visual components were filmed, edited, and manipulated by Wentworth.

To accomplish the objective of affecting narrative through diegetic, extradiegetic, and metadiegetic motifs [1], examples were collected and explored from multimodal multimedia materials. Additionally, the manipulation of spatiotemporal experience and the evocation of memory as a representation of literal and metaphoric audio-visual events were also investigated. This research was used to inform the creative practices applied in the final work.

Within this paper, we discuss the techniques applied in the use of geographic space for the manipulation of personal memories of place from the perspective of a multimedia arts project. Furthermore, discourse is presented on performance observations to support the project theories on the narration of place. The function of these observations was to informally explore the concepts applied in the composition, to examine the philosophies in summoning narrative from within the observer, and finally to evaluate the attempts made to cross boundaries between the audiences' imagination and reality.

## 2. BACKGROUND

The phenomenological theory of art highlights that the aesthetic response to art is as equally important as the art itself, a concept explored in reader – response literature [2]. In the context of a multimedia performance, aesthetic realisation can therefore only be accomplished by the audience and artistic appreciation can only be found in the performance of the work itself. Performances are consequently only to be considered more than the presentation of multimedia once they have been interpreted by the audience. It was therefore hypothesised that an artistic multimedia performance can be conceived of to engage with an audiences' imagination by leaving room for self-interpretation; by allowing the audience to create their own narrative place from the presented spatial materials.

Concepts of space and place, as contrasting entities, have been studied for many years [3]. These theories have since been extended to define the concept of space as abstract, and that place refers only to specific environments that are invested with emotional value [4]. Other researchers have further separated concepts of space and place [5], where space can be used absolutely to define the structure of the world and the three-dimensional environment in which we inhabit, and that place is formed from the cultural understandings attached to them.

In contrast, narrative space is defined as the place or places within which situations and events are represented and narrating instances occur [6]. In this case, a narrative space serves no other function but to supply the setting of the narrative. However, narrative space involves significantly more than descriptions and references to landmarks associated with the conventional scene-setting accounts; as is seen in other narrative theories and reader - response studies. This serves to highlight that "narrative setting" cannot be separated out from the "narrative voice" or that geographical space can be treated as a unique place without human interpretation [7]. In narrative studies, examinations of the construction of time and space as orientation elements have been discussed and serve as a reference for the analysis and interpretation of narratives [8].

In terms of arts practice, if an audience is presented with field recordings of water flowing, it is supposed that they can imagine a babbling brook; one from their own memories or previous experiences perhaps. However, if the audience is also presented with video footage of a river, then they can no longer imagine this scene, as the imagination is presupposed by the image. The difficulty in achieving the outlined project goals was therefore to present the audience with enough information to evoke memories, while also informing them of actual knowledge from within the orchestrated media. The unified whole or gestalt of the performance must not therefore be presented as having literal meaning, but it must instead be constructed as something that is intangible or inconsequential.

In this context, the "U - Modified" project was to deal with the construction of space and place in a narrative that served to evoke previous accounts of movement from one space to another. However, the focus was not to understand how listeners engage with interactive works, but to try to manipulate their understanding of narrative events and present different settings for narrative place.

## 3. PRECEDING WORK

It was first observed that live performance of "U" would guide an audience on an internalised journey by revealing a varied sound structure that extended, imitated, developed, and distorted sonorities from the real-world [9]. Throughout this composition, intangible sonic patterns were displayed that carefully guided and stimulated the audience to reflect upon previous experiences of place through sound. By recalling memories of reality and appealing to previous experiences of place, a spatiotemporal connection was created through the associated memories. This presented a journey for the individual audience members, through literal space and internalised place, one that was intended to be unclear and unresolved throughout. Vagaries between realism, recollection, and unconscious thoughts were intentionally produced to further confound these experiences.

Following an initial analysis of this composition, visual elements were collected and recorded by Wentworth, in directed answer to her own personal associative and visceral responses to the soundscapes of "U". To Wentworth, the sonic landscape felt fragmented, sporadic, and somewhat uneasy, and recalled travel, memory of place, and daydreams. Therefore, in retort, source footage was taken from multiple locations around Europe as well as footage from transit between locations. This included themes of naturally occurring elements, such as sky, clouds, wind, and water, as well as transportation by land, sea, and air.

The collected footage was then composited via programming in Max/MPS/Jitter. The patch applied waveform amplitude analyses of the individual 5.1 audio channels to correspond the opacities of several overlaid video clips. This further fragmented and juxtaposed the video content materials, as multiple geographical locations flickered in and out according to the dynamic level of the separate audio-channels. These familiar spaces, in combination with the soundscapes of "U", were constructed and arranged to evoke the fragmented feeling one gets when frequently travelling between borders and changing geographic location.

The combined composition was therefore to exist as a symbolic space and an imaginary place, one that was evocative, yet still abstract to the audience. All of these effects were to be achieved by simultaneously presenting two important factors. Firstly, the presentation of traditionally recognisable multimedia samples with room-noise (the kind that is ever present in conventional room settings). Secondly, to conjure within the audience questions about the significance of the media and how they relate to the audiences' own personal identities. These themes were supported within the composition through careful editing and arrangement of tape, vocals, and video clips, their motifs serving to further blur the reality of space with interpretations of place.

## 4. INFLUENCES

Within the final composition, the sonic arrangements are deeply influenced by Pierre Schaefer and his work on the theoretical basis of Musique Concrète as a compositional practice [10]. This is demonstrated in the recording, manipulation, and arrangement of audio samples from multiple sources. This philosophy was then also applied to the visual measures of the piece. Specifically, influential were Schaefer's use of voice, as demonstrated in Scherzo [11] and Apostrophe [12]. Also, of influence were the compositions of Luc Ferrari. Specifically, Ferrari's use of the superposition of cycles in the creation of new events and the application of environmental sounds; as heard in Tautologies III [13]. Furthermore, Ferrari's work was also influential in the crafting of found-sounds and the use of casual narratives, like those Ferrari demonstrated in Anecdotiques [14], Petite symphonie intuitive pour un paysage de printemps [15], and Heterozygote [16].

Other, less distinctive influences can also be identified from the artistic and compositional styles from: Karlheinz Stockhausen's use of silence and atmosphere in Telemusik, Studie II, and Kontakete; Luciano Berio's use of vocal sounds and insanity narratives in Visage; the use of breakneck editing, stabs, jabs, long drones, and echoes from Bernard Parmegiani's De Natura Sonorum and Hors Phase; Electronique and the provoked audience reactions in Déserts by Edgar Varése; the dream like sounds, narratives, and use of diegetics in Automatic Writing by Robert Ashley; in the mixing of Elektronische Musik with Musique Concrète as heard in Morton Subotnick's Silver Apples of the Moon; John Chowning's Stria; John Cage's Imaginary Landscape; as well as the film score influences of David Lynch.

By exploring previous works in this context, it was observed that the human psyche possesses the ability to exist as both an exterior and interior and that these settings can often present themselves in a somewhat contradictory way. Therefore, with the intentional confusion of what is inside from what is outside, the performance of "U - Modified" was to embody the role of the uncanny in its narrative. That is, the opposite of what is familiar to the listener's usual internalised narrative [17]. In doing so, the composition presented a challenge to the audience. The individual audience members were to confront that which was explicitly expressed in the media presented before them and in response create reality, truth, and self-narration within. "U-Modified" therefore deliberately created a blurring between the spaces that the audience were seeing and hearing with respect to their own past experiences of place.

## 5. PERFORMANCE

For the performance of "U-Modified", the audiences' experiences were to be distracted from the reality of the physical concert space by being presented with multimodal stimulation while stationary, silent, and seated in a dimly lit room. The presentation of the work in this manner imposed the role of the internal narrative invocation, highlighting the unconscious desire for external stimulation and the discomfort that arises from the removal or dulling of the Aristotelian senses. This performance requirement emphasised how the listening environment and delivery methodology was to play an important role in presenting the audience with the opportunity to investigate the existence of a self-constructed internalised alternative-reality and the systematic temporalities that present themselves therein. This also served to underscore the intrinsic authority of the unconscious mind over the physical perception of place.

During the performance, the collected video samples of familiar occurrences, were presented to the audience on a large projection screen. The video footage was then abstracted throughout in relation to the amplitudes of the created soundscape, as described earlier. Throughout the composition, sound and image, with recognisable source and space, were presented to the audience and made accessible with obviously explicit diegetic meaning. Memories of space the audience had previously experienced were therefore evoked and set against the backdrop of the performance. Consequently, for the narration and creation of a world in which these events were to take place and for the audience to apply these in the telling of a perceptibly ambiguous story of geographic spaces a previous memory or knowledge was required. However, the evoked memory could not be accurately realised by the presented spaces, as this would require the audience members to have visited the location previously; where memory and perception would then be in agreeance.

The sources of these recognisable events therefore presented themselves to the audience, evoking previous memories or instances recounted from their own reality. Furthermore, there existed within the performance extradiegetic media, as sound and image, that passed directly from the collected materials to the observer, unmediated by any specific narrative tool or primary narrative source. These events were to serve the audience directly and invoke feelings or moods within them.

The blending of diegetic sound with extradiegetic was applied with the purpose of further blurring reality with abstract thoughts; aspects of the internalised narrative of the listener. These diegetic sounds, within the overall composition, were to appear unnaturally amplified and distorted. Naturally occurring sounds and recognisable images were mutated and mixed with mutable silences and blackouts to create audio-visual discomfort in the observer. In achieving this, the recognisable sources were intertwined with darkness, silence, hum, and noise. The latter of these, served to interrupt the imaginary narrative as natural silences, the silence that one experiences between events in natural everyday environments.

## 6. ARRANGEMENT

The performance methodologies applied within the piece were not intended to act as a direct representation of an idea of place or present an explicit narrative, nor were they to exist on a purely aesthetic platform either. The requirement for audience participation, through recollection, emotional reaction, and contextual awareness was to operate as an expressive outlet for the channelling of the personal significance of place from space. These meanings were not to be innate to the composition, but they were to be evoked into existence by the audience; engendering an emotional space with which there was no literal foundation. The reality of the spaces presented were fundamentally plausible and realistic, but they also conjured from memory and imagination formed from previously experienced places.

As the arrangement drifts between periods of darkness, images, and sounds of everyday spaces the audio and visual elements combined and transitioned from external stimulation to the internal imagination, affecting the audience's intrinsic relationships to extrinsic influences. More importantly, it was to affect the audience's relationship with real-time. To hold the audience captivated in a continuous flow, would be natural for some level of immersion. However, a sequence of continuous narrative removed all anticipation of variation.

With the linear nature of time, specifically in immersive soundscape environments, time can be destabilised through total immersion. The subjectivity of the listening experience must capture the audience and manipulate their internal spatiotemporal relationship with the real-world. In this, the performance allowed for the combining of past with present and to form the narrative space. However, the expectation of the audience was altered to incorporate unexpected twists and to purposefully frustrate the expected outcomes.

## 7. REFLECTIONS ON PERFORMANCES

The observed performances of "U-Modified", in the traditionally obvious diegetic presentation of collected materials, over-emphasised the importance of silence in day-to-day real-world interactions. Equally, the conventional role of the diegetic narrative was demoted to the background. The treatment of stillness within the piece altered the role of silence, that is, in the absence and abundance of stimulation that normally indicates a presence of, or lack of, importance.

Through accomplishing this, the audiences' assumptions were clearly challenged. That the sound and imagery in the foreground were significantly more important than those in the background. Furthermore, the relative intensity of a spatial object did not necessarily increase its worth or meaning. For example, traditionally, it is understood that in an intimate conversation in a crowded room, the speaker directly in front of the listener is more important than the other speakers and conversations happening simultaneously around them. In challenging the accepted relationships between sound and vision, the relative correlation between the two, the virtual embodiment of a narrative within the audience, and the significance of a stimulus within the narrative served to further engage the audience, invigorating these dissonances and thrusting forth the uncanny.

Within the piece, challenges to the role of extradiegetic music were also presented. Short musical phrases, and their repetition as leitmotifs, were purposefully included in the score. However, they were employed unconventionally to raise ambiguous reference to their own existence, as opposed to the signifier of obvious narrative events. These stingers and leitmotifs were external to the narrative, unattached to the more obvious diegesis. However, they were also applied as indicators and markers that referred to the narrative tools that they were accompanying; that is, those that they first appeared with. Their application was, therefore, intentionally confused due to their non-linear application. They occurred throughout but referenced each other and the brief moments in which they previously occurred.

By cross-referencing these methodologies as they occurred also further compounded their effect. In addition, sustained chords, loops, and drones were applied in such a way that they denoted no apparent theme or motif that had previously occurred in a deceptive or contradictory state. Melodies and motifs appear and disappear, evoked a response, and then moved on. Never having the same meaning twice allowed them to pass through their individual significance with regards to time, space, and reality.

Disorientation in the narrative space and the physical act of storytelling also has consequences on the interplay between time and space in the association of the narrative place. Through the application and exploration of these factors, attention was drawn to dimensions of both the familiar and unfamiliar in the creation of an audiences' own narrative. This was achieved through the reversal of conventional understandings of the diegetic, the importance of sound through elusive changes in volume, to raise doubt and question the existence of an imagined source embodied within the listener, and by modifying the role of the extradiegetic features within the internal narrative.

The overlapping of diegetic with extradiegetic motifs presented within real-world environments further augmented the challenges of addressing the research projects objectives. The presentation of space within the piece promoted a blurring of personal experience and the sonic representation of narrative. The personification of dreams and reality within a predetermined space served to evoke a site-specific inner narrative within the listener and create indistinguishable borders between the two. The extradiegetic presentation of room-noise was constant, and representative of the silences and everyday experiences mentioned earlier.

Background noises buzzed, hissed, and whirred about the audience continuously. It was unremittingly present and ignored the traditional rules that govern applications in extradiegetic scoring. The uncanny nature of these sounds raised questions relating to its abandon of borders; in both literal and imaginary contexts. Amplification above the "normal" added to the unnerving nature of its existence; dreams are never that loud and reality is never that noisy. Therefore, the appearance of an extradiegetic motif within the soundscape presented itself as a diegetic instance and vice versa.

The inclusion of transdiegetic movements, the crossing over of both the diegetic and non-diegetic, created evolution and gathered meaning from the subjective interpretation of the imaginary narrative created by the listener. Because of this, there also existed metadiegetic narratives, a secondary narrative embedded within the primary narrative, which was dependent upon the internal functioning of the media in relation to its interpretation by the listener. This further developed the dream/reality and space/place metaphor by articulating movement through narrative levels.

This movement also spilled further beyond that of the presented narrative, consuming the physical extradiegetic space of the listener. It was observed that the fictional space within the performance space was to encroach upon the physical space within which the audience resided. In this, the meaning of place was used to apprehend the unfamiliar spaces that the audience were presented with.

The leitmotifs and their function within this composition were not tied directly to any one disposition or circumstance; they were applied in such a manner to drift between diegetic, extradiegetic, and metadiegetic motifs alike. The boundaries between the narration, perception, and experience of the listener were therefore pushed beyond physical space and into a transdiegetic liminal place.

## 8. CONCLUSION

In conclusion, the samples and materials that were collected and edited by the project group were purposeful and significant in meaning to the overall objectives and motivation of the piece. It was observed that the timbre of the sounds produced within the composition and their origins could draw fascination from the audience. Furthermore, with the removal of measure and arithmetic progression from the piece, the audience were left only the timbral qualities of the untreated sound and augmented samples.

In addition, the movement of narrative through diegetic, extradiegetic, and metadiegetic, the manipulation of the spatiotemporal experience, and the evocation of memory were all represented through actual and metaphorical sonic events. The movement of trees in the wind, the blowing of pipes, industrial machinery in motion, and electrical short-circuits represented actions that the listener could associate with familiar timbres of movement.

This familiarity was further manipulated with room-tone and sounds of a low rumbling nature, timbres that may not be consciously processed, but existed in common places. The timbres of these sounds were indeterminately presented both delicacy and violence but were always ethereal. With tone, expression, and a unique inflection being ever present. In observing reproductions of this work, when formulating opinions of the performance, the audience was acutely aware of these factors immediately post-performance and perhaps even further in retrospect.